\documentstyle[epsfig]{l-aa}

\def\Msun{M$\sb{\odot}$}
\def\Lsun{L$\sb{\odot}$}

\def\Mbol{$M\sb{\rm bol}$}

\def\kms{~km~s$\sp{-1}$}

\begin{document}
\thesaurus{ 6(02.01.2 -- 08.02.4 -- 08.02.5 -- 08.09.2: HR 363, HD
191226) }
\title{A CORAVEL radial-velocity monitoring of S stars: 
symbiotic activity vs. orbital
separation (III)\thanks{Based 
on observations performed with the Swiss 
telescope at the Haute-Provence Observatory, France}}
\author{J.M. Carquillat\inst{1}
\and
A. Jorissen\inst{2}\thanks{Research Associate, F.N.R.S., Belgium}
\and
S. Udry\inst{3}
\and 
N. Ginestet\inst{1}
}

\offprints{J.M. Carquillat}
\institute{
Observatoire Midi-Pyr\'en\'ees, UMR 5572, 
Av. E. Belin 14, F-31400 Toulouse,
France
\and 
Institut d'Astronomie et d'Astrophysique, Universit\'e Libre de
Bruxelles,
Campus Plaine C.P. 226,  Bd du Triomphe, B-1050 Bruxelles, Belgium
\and
Observatoire de Gen\`eve, CH-1290 Sauverny, Switzerland
}
\date{ Received date; accepted date }
\maketitle
\markboth{J.M. Carquillat et al.: CORAVEL orbits of two S stars} 
{J.M. Carquillat et al.: CORAVEL orbits of two S stars} 

\begin {abstract}
Orbital elements are presented for the Tc-poor S stars HR 363 (= HD
7351) and HD 191226. With an orbital period of 4592~d (=12.6 y), HR
363 has the longest period known among S stars, and yet it is a
strong X-ray source. Its X-ray flux is similar to that of HD 35155,
an
S star with one of the shortest orbital periods (640~d). This
surprising result is put in perspective with other
diagnostics of binary interaction observed in binary S stars. They
reveal that there is no correlation between the level of binary
interaction
and the orbital period. All these activity diagnostics moreover
exhibit a strong time-variability. 
In the well-documented case of HR 1105, this time-variability
appears
to be a combination of orbital modulation and  secular variation.
A stream of gas from the red-giant wind, which is heated when
funneled
through the inner Lagrangian point, has been proposed as the source
of
the hard photons (Shcherbakov \& Tuominen 1992). 
Different viewing angles of the stream during the orbital cycle
account for
the orbital modulation, whereas long-term fluctuations of the
mass-loss rate account for the secular variations. Little
dependence to the orbital separation is expected for this kind  
of activity. If such streams are causing the activity observed in
the
other binary S stars as well, it would provide a natural
explanation
for the absence of correlation between orbital periods and activity
levels, since the red-giant mass loss rate would be the dominant
factor. The existence of such funneled streams is moreover
predicted 
by {\it smooth particle hydrodynamics} simulations of mass transfer
in
detached binary systems.
\keywords{stars: S -- stars: HR 363, HD 191226 --
binaries: spectroscopic -- binaries: symbiotic -- accretion}
\end{abstract}

\section{Introduction}

This paper is the third one (see Udry et al. 1998ab for the first two) in a 
series presenting new orbits obtained for barium and S stars as a
result of a
decade-long monitoring with the spectrovelocimeter CORAVEL (Baranne
et
al. 1979)  installed on the 1-m Swiss telescope at the 
Haute-Provence Observatory (OHP, France).
 
The two S stars HR 363 (=HD 7351) and HD 191226, that are the
subject
of this paper, were put on the CORAVEL monitoring program by both 
the Toulouse and the Geneva-Brussels teams, as part of larger
samples. 
The Toulouse team aims at obtaining orbital elements for a sample of
bright late-type stars quoted in the literature as being spectroscopic binaries, 
but with no orbit available. 
For the Geneva-Brussels team, HD 191226 and HR 363
belong to a sample of Tc-poor S stars, a family of
chemically-peculiar red giants suspected of being all binaries. 
A detailed discussion of the `binary paradigm' for Tc-poor S stars
can
be found in Jorissen et al. (1998), and needs not be repeated here.

The new orbits of HD~191226 and HR~363 are provided in
Sects.~\ref{Sect:HD191226} and \ref{Sect:HR363}, respectively. 
Diagnostics of binary interaction are available for these two systems
from published observations in the UV and X domains.
The availability of their orbital
elements now offers the possibility to correlate these diagnostics of
binary interaction with the orbital separation, as discussed 
in Sect.~\ref{Sect:UV}.

\section {HD 191226}
\label{Sect:HD191226}

The spectroscopic binary nature of HD 191226 is already mentioned
in
the {\it General Catalogue of Stellar Radial Velocities} (GCSRV;
Wilson 1953): from 10 measurements obtained at the Mount Wilson
and Victoria observatories, the radial velocity of this gM2 star 
varied by 22 \kms. Also classified as M2III by Nassau \& McRae
(1949),
whereas Barbier (1962) assigns it the spectral type K2II:, HD
191226
is finally recognized by Keenan \& Boeshaar (1980) as being a weak
S star
(M1S-M3SIIIa). The spectral class MxS refers to stars with the
strongest ZrO bands barely visible. A spectrum obtained in the near
infrared at a dispersion of 3.3 nm/mm with the Aur\'elie
spectrograph
on the 1.52-m telescope at OHP  
yields a M0-M1III+ classification from a comparison with spectral
standards (Ginestet et al. 1994; Carquillat et al. 1997). 
This spectrum is unfortunately of no
use to confirm the S-type classification of HD 191226, as the
typical S
spectral features are better seen in the optical domain (Jaschek \&
Jaschek 1987). The Tc-poor nature of HD 191226 was reported by
Little
et al. (1987) and Smith \& Lambert (1988).

\renewcommand{\baselinestretch}{1.}
\begin{table}
\caption[]{\label{Tab:VrHD191226}
Radial velocities of HD 191226. Column 4 ($\epsilon_1$) lists the
uncertainty on the individual measurements. The columns labelled
`Phase' and `O-C'
refer to the orbital solution listed in Table~\ref{Tab:orbit}. The
column
labelled `Obs' provides the origin of the radial velocity: COR =
CORAVEL; Cou
= Coud\'e spectrum obtained at OHP; McD = data obtained by Brown et
al. (1990) at McDonald observatory
}
\begin{tabular}{lllccllllll}
\hline
 HJD       & Phase & \multicolumn{1}{c}{RV}  
                   & \multicolumn{1}{c}{$\epsilon_1$} 
                   & \multicolumn{1}{c}{$O-C$} 
                   & Obs \cr
2\ts400\ts000+&    & \multicolumn{1}{c}{(\kms)}
                   & \multicolumn{1}{c}{(\kms)}    
                   & \multicolumn{1}{c}{(\kms)} 
                   & \cr
\hline
 46337.343& 0.229& $-$22.68 &   0.32 & 0.0&COR \cr
 46587.548& 0.436& $-$21.16 &   0.30 & 0.0&COR \cr
 46719.370& 0.544& $-$22.20 &   0.34 &$-$0.1&COR \cr
 46965.534& 0.748& $-$26.68 &   0.48 &$-$0.4&COR \cr
 47099.324& 0.858& $-$28.86 &   0.36 &+0.4&COR \cr
 47100.360& 0.859& $-$29.94 &   0.40 &$-$0.6&COR \cr
 47285.598& 0.012& $-$29.86 &   0.30 &$-$0.1&COR \cr
 47343.000& 0.060& $-$28.7  &   0.7  &$-$0.4&McD \cr
 47369.000& 0.081& $-$28.6  &   0.7  &$-$1.1&McD \cr
 47397.462& 0.105& $-$26.0  &   0.9  &+0.6&Cou \cr
 47460.291& 0.157& $-$24.81 &   0.33 &$-$0.1&COR \cr
 47497.000& 0.187& $-$24.5  &   0.7  &$-$0.8&McD \cr
 47519.000& 0.205& $-$22.7  &   0.7  &+0.5&McD \cr
 47868.264& 0.494& $-$22.15 &   0.30 &$-$0.5&COR \cr
 48128.435& 0.709& $-$24.64 &   0.32 &+0.7&COR \cr
 48137.444& 0.716& $-$25.06 &   0.29 &+0.4&COR \cr
 48936.362& 0.376& $-$21.54 &   0.32 &$-$0.4&COR \cr
 49100.589& 0.512& $-$21.77 &   0.29 & 0.0&COR \cr
 49145.572& 0.549& $-$22.65 &   0.31 &$-$0.4&COR \cr
 49185.575& 0.582& $-$23.39 &   0.30 &$-$0.7&COR \cr
 49250.322& 0.636& $-$24.18 &   0.29 &$-$0.6&COR \cr
 49319.277& 0.693& $-$25.15 &   0.29 &$-$0.2&COR \cr
 49482.565& 0.827& $-$28.48 &   0.31 & 0.0&COR \cr
 49546.501& 0.880& $-$29.81 &   0.31 & 0.0&COR \cr
 49639.309& 0.957& $-$30.47 &   0.29 &+0.1&COR \cr
 49640.402& 0.958& $-$31.03 &   0.30 &$-$0.4&COR \cr
 49643.329& 0.960& $-$30.24 &   0.29 &+0.4&COR \cr
 49783.716& 0.076& $-$27.37 &   0.32 &+0.3&COR \cr
 49899.537& 0.172& $-$24.31 &   0.30 &$-$0.1&COR \cr
 49959.354& 0.221& $-$22.45 &   0.33 &+0.4&COR \cr
 50194.611& 0.416& $-$21.08 &   0.31 & 0.0&COR \cr
 50314.419& 0.515& $-$21.33 &   0.29 &+0.5&COR \cr
 50329.496& 0.527& $-$21.63 &   0.30 &+0.3&COR \cr
 50359.335& 0.552& $-$21.42 &   0.29 &+0.8&COR \cr
 50382.257& 0.571& $-$22.21 &   0.31 &+0.3&COR \cr
 50416.295& 0.599& $-$22.81 &   0.40 &+0.1&COR \cr
\hline
\end{tabular}
\end{table}

\begin{table}
\caption[]{\label{Tab:orbit}
Orbital elements of HD 191226 and HR 363
}
\begin{tabular}{ll@{$\pm$}ll@{$\pm$}llllllll}
\hline
        & \multicolumn{2}{c}{HD 191226}  & \multicolumn{2}{c}{HR
363} \cr
\hline  
$P$ (d)             & 1210.4        & 4.3    & 4592.7        &
110.0\cr
$T_0$               & 2\ts448\ts481 & 24     & 2\ts444\ts696 & 230 
\cr
$e$                 & 0.19          & 0.02   & 0.17          & 0.03
\cr
$V_0$ (\kms)        & $-$25.05        & 0.08   & 1.55          & 0.15
\cr
$\omega$ ($^\circ$) & 207.4         & 7.5    & 104.3         & 13.7
\cr
$K$ (\kms)          & 4.76          & 0.11   & 5.43          & 0.16
\cr
$A_1 \sin i$ (Gm)   & 77.76         & 1.85   & 337.5         & 12.9
\cr
$f(M)$ (\Msun)      & 0.0128        & 0.0009 & 0.073         &
0.007\cr
$\sigma(O-C)$ (\kms)&
\multicolumn{2}{c}{0.38}&\multicolumn{2}{c}{0.68}\cr
$N$                 & \multicolumn{2}{c}{36} 
&\multicolumn{2}{c}{50}&\cr 
\hline
\end{tabular}
\end{table}

\begin{figure}
  \vspace{9cm}
  \vskip -0cm
  \begin{picture}(8,8.5)
    \epsfysize=9.8cm
    \epsfxsize=8.5cm
    \epsfbox{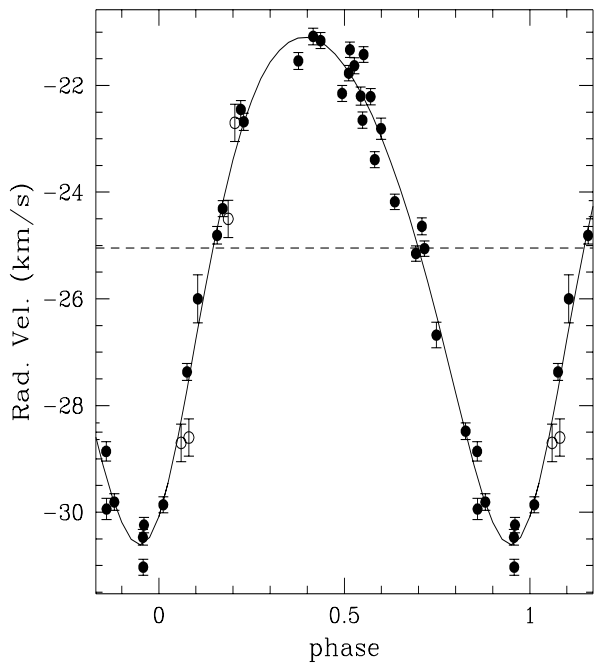}
  \end{picture}
  \vskip -0.5cm
\caption[]{\label{Fig:VrHD191226}
The radial-velocity curve of HD 191226, folded with the orbital
period. Filled dots refer to OHP data (CORAVEL and Coud\'e), and open 
dots to McDonald data
}
\end{figure}

The orbital elements of HD~191226 have been derived from the 36
radial-velocity  measurements listed in Table~\ref{Tab:VrHD191226}.
Among these, 31 were obtained with the CORAVEL spectrovelocimeter
at
OHP (with an average uncertainty of $\bar{\epsilon} = 0.34 \pm
0.06$
\kms). One has been obtained at the Coud\'e focus of the 1.52-m
telescope at OHP, on a baked IIaO plate (\#\ GA 8139), with a 2.0
nm/mm
dispersion. Four radial-velocity measurements obtained by Brown et
al. (1990) on the 2.1-m telescope at McDonald Observatory equipped
with a Reticon detector have also been used in the orbital
solution. 
All these radial velocities, covering 3.37 orbital cycles, are on
the  
system of IAU standards.  The orbital
elements listed in Table~\ref{Tab:orbit}
have been computed with the BS1 program (Nadal et al. 1979)
by assigning a weight 1 to the CORAVEL data and 0.25 to the five
other
measurements. The radial-velocity curve, folded with the orbital
period, is presented in Fig.~\ref{Fig:VrHD191226}.
Older data, obtained through the years
1912--1934, were not used in the orbital solution, because their
accuracy is not good enough, even to improve the orbital period.
The Mount Wilson data have (when available) uncertainties of the
order
of 2.5 to 3.3 \kms, and their variation range ($-13$ to 35 \kms;
Abt
1973) appears incompatible
with the small value of the semi-amplitude ($K = 4.8$~\kms;
Table~\ref{Tab:orbit}) derived from modern, more
accurate measurements. As far as the the radial velocities obtained
at
Victoria by Harper (1934) are concerned,
they are of doubtful quality (both plates are qualified as `weak')
and
were therefore not used either.

An estimate of the mass of the companion can be obtained from the 
mass function listed in Table~\ref{Tab:orbit}.
Adopting a  mass of $1.5\pm0.5$~\Msun\ for the S star, which seems
reasonable given its M2III spectral type (e.g., Schmidt-Kaler 1982;
see also Jorissen et al. 1998),    
a minimum mass of $0.35\pm0.07$ \Msun\ for the secondary follows 
from
the condition $\sin i \le 1$. Taking $M_1 = 1.5$ \Msun\ for the
red
giant and $M_2 = 0.6$ \Msun\  for its companion yields $i =
40^\circ$,
$q = M_1/M_2 = 0.4$ and $A_1 = 121$ Gm, or in terms of orbital
separation, $A = A_1 + A_2 = A_1 (1 + 1/q) = 2.8$ AU. 

The companion mass is thus compatible with that of a white dwarf
(WD).
Observations in the UV with the {\it Goddard High Resolution
Spectrograph} (GHRS) on board the 
{\it Hubble Space Telescope} indeed reveal the presence of a WD
with  $T
\sim 15\ts000$ K (Ake 1997). These
results will be discussed further in Sect.~\ref{Sect:UV}.

\section{HR 363}
\label{Sect:HR363}

As for HD~191226, the spectroscopic-binary nature of HR~363 (=
HD~7351) is already mentioned in the GCSRV, where it is classified
as gM2. Six radial-velocity measurements are mentioned in that
catalogue, among which 4 were performed at the David Dunlap
Observatory (Young 1945) and 2 at the Mount Wilson Observatory
(Wilson \& Joy 1952; Abt 1970). These two sets of measurements
yield average velocities of +5.8 and $-1.8$~\kms, respectively. It
is likely that it is this difference which led to suspect the binary
nature of HR~363 in the GCSRV.
More recently, three radial-velocity measurements were obtained at
the E.W. Fick Observatory (Beavers \& Eitter 1986), and 15 more by Brown
et al. (1990). The latter measurements, covering 1000~d, confirmed
the spectroscopic-binary nature of HR~363, and made it clear that
its period was quite long. 

HR 363 was classified M2S by Keenan (1954), who noted its strong
BaII $\lambda 455.4$~nm line, and M3IIS by Yamashita (1967). Keenan \&
Boeshaar (1980) reclassified it later on as S3+/2- in their revised
classification scheme (where the first digit is a temperature
index, and the second one is an index of ZrO strength, 2
corresponding to ZrO $<$ TiO). More recently, Sato \& Kuji (1990)
have classified HR~363 as M2III. These authors stress that,
although HR~363 is often considered as an S star, its salient
spectral features are those typical of M giants, except for a
strong BaII $\lambda 455.4$~nm line. 
As for HD~191226, a near-infrared spectrum has been obtained for
HR~363 with the Aur\'elie spectrograph at OHP, and yields a M3III
spectral type. 
Finally, regarding its Tc-poor
nature, we refer to Little et al. (1987) and Smith \& Lambert
(1988).       

\tabcolsep 3pt
\begin{table}
\caption[]{\label{Tab:VrHR363}
Radial velocities of HR 363 = HD 7351. Symbols are as in Table~1
}
\begin{tabular}{llcccllllll}
\hline
 HJD       & Phase & \multicolumn{1}{c}{RV}  
                   & \multicolumn{1}{c}{$\epsilon_1$} 
                   & \multicolumn{1}{c}{$O-C$} 
                   & Obs \cr
2\ts400\ts000+ &   & \multicolumn{1}{c}{(\kms)}
                   & \multicolumn{1}{c}{(\kms)}    
                   & \multicolumn{1}{c}{(\kms)} 
                   & \cr
\hline
 43026.6290& 0.637 &  +6.0   &   0.8  & +0.5&Cou\cr
 45340.3280& 0.140 &  $-$3.73  &   0.32 & +0.4&COR\cr
 45343.2657& 0.141 &  $-$4.21  &   0.33 & $-$0.1&COR\cr
 45595.6070& 0.196 &  $-$4.62  &   0.31 & $-$0.7&COR\cr
 46008.4922& 0.286 &  $-$1.86  &   0.31 & +0.6&COR\cr
 46015.4520& 0.287 &  $-$1.98  &   0.30 & +0.5&COR\cr
 46296.5681& 0.349 &  $-$1.65  &   0.34 & $-$0.5&COR\cr
 46335.5978& 0.357 &  $-$0.56  &   0.31 & +0.3&COR\cr
 46715.5295& 0.440 &  +0.87  &   0.33 & $-$0.3&COR\cr
 46722.5150& 0.441 &  +2.10  &   0.32 & +0.9&COR\cr
 47097.4289& 0.523 &  +2.23  &   0.34 & $-$1.0&COR\cr
 47101.4738& 0.524 &  +2.90  &   0.31 & $-$0.3&COR\cr
 47459.4887& 0.602 &  +4.20  &   0.31 & $-$0.7&COR\cr
 47463.4908& 0.603 &  +4.96  &   0.30 & +0.1&COR\cr
 47867.3641& 0.691 &  +6.42  &   0.31 & +0.1&COR\cr
 47872.3239& 0.692 &  +6.28  &   0.30 & +0.0&COR\cr
 48128.5884& 0.747 &  +7.10  &   0.30 & +0.4&COR\cr
 48261.3514& 0.776 &  +7.58  &   0.38 & +0.9&COR\cr
 48671.2717& 0.866 &  +5.68  &   0.32 & +0.2&COR\cr
 48676.2604& 0.867 &  +5.77  &   0.31 & +0.3&COR\cr
 48936.5273& 0.923 &  +2.36  &   0.30 & $-$1.2&COR\cr
 48967.3589& 0.930 &  +2.65  &   0.30 & $-$0.7&COR\cr
 48972.4352& 0.931 &  +2.68  &   0.34 & $-$0.5&COR\cr
 49002.2925& 0.938 &  +2.51  &   0.30 & $-$0.5&COR\cr
 49313.4231& 0.005 &  +0.15  &   0.30 & +0.5&COR\cr
 49317.3760& 0.006 &  +0.75  &   0.29 & +1.0&COR\cr
 49321.4032& 0.007 &  +0.54  &   0.34 & +0.8&COR\cr
 49371.2581& 0.018 &  +0.21  &   0.31 & +1.1&COR\cr
 49639.6748& 0.076 &  $-$3.09  &   0.28 & $-$0.1&COR\cr
 49640.5580& 0.077 &  $-$2.98  &   0.30 & +0.0&COR\cr
 49730.2663& 0.096 &  $-$3.54  &   0.28 & +0.0&COR\cr
 49734.2638& 0.097 &  $-$4.01  &   0.27 & $-$0.5&COR\cr
 49781.2730& 0.107 &  $-$4.54  &   0.38 & $-$0.8&COR\cr
 49783.2658& 0.108 &  $-$4.70  &   0.32 & $-$1.0&COR\cr
 49785.2729& 0.108 &  $-$4.57  &   0.35 & $-$0.9&COR\cr
 49964.6007& 0.147 &  $-$4.08  &   0.31 & +0.0&COR\cr
 50042.3732& 0.164 &  $-$3.61  &   0.30 & +0.5&COR\cr
 50052.3666& 0.166 &  $-$3.55  &   0.29 & +0.6&COR\cr
 50072.2763& 0.171 &  $-$3.89  &   0.31 & +0.2&COR\cr
 50083.3657& 0.173 &  $-$3.18  &   0.32 & +0.9&COR\cr
 50123.2461& 0.182 &  $-$3.41  &   0.30 & +0.6&COR\cr
 50325.5841& 0.226 &  $-$3.75  &   0.31 & $-$0.2&COR\cr
 50355.4486& 0.232 &  $-$4.29  &   0.30 & $-$0.8&COR\cr
 50363.4453& 0.234 &  $-$5.20  &   0.29 & $-$1.7&COR\cr
 50382.3557& 0.238 &  $-$4.31  &   0.30 & $-$0.9&COR\cr
 50415.4728& 0.245 &  $-$2.81  &   0.29 & +0.5&COR\cr
 50420.3984& 0.246 &  $-$2.59  &   0.31 & +0.7&COR\cr
 50428.3299& 0.248 &  $-$3.79  &   0.30 & $-$0.6&COR\cr
 50476.3083& 0.259 &  $-$1.70  &   0.30 & +1.3&COR\cr
 50615.3080& 0.289 &  $-$2.01  &   0.30 & +0.4&COR\cr
\hline
\end{tabular}
\end{table}

The orbital  solution for HR 363 listed in Table~\ref{Tab:orbit} 
is based on 49 CORAVEL radial-velocity
measurements covering 1.1 orbital cycle (from 1983 to 1997), to
which one older measurement obtained in 1976  has been added   
(Table~\ref{Tab:VrHR363}). This early measurement (obtained at the
1.52-m telescope of Haute-Provence Observatory on the photographic
spectrum GA 2881) substantially
improves the period determination, since it brings the orbital
coverage to 1.65 cycle. Its accuracy is only 0.8~\kms, compared to
an average of 0.30~\kms\ for the CORAVEL measurements. It
has therefore been attributed a weight of 0.25 in the orbital solution
(compared to 1 for the CORAVEL measurements).

\begin{figure}
  \vspace{9cm}
  \vskip -0cm
  \begin{picture}(8,8.5)
    \epsfysize=9.8cm
    \epsfxsize=8.5cm
    \epsfbox{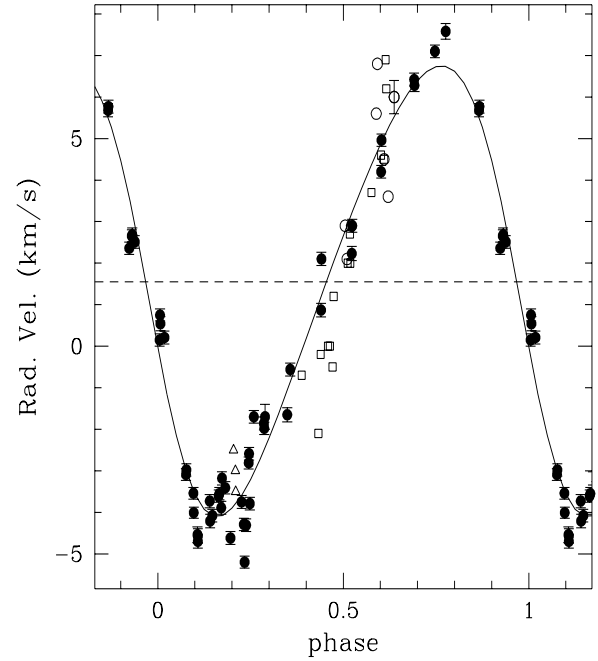}
  \end{picture}
  \vskip -0.5cm
\caption[]{\label{Fig:VrHR363}
The radial-velocity curve of HR 363 (=HD 7351), folded with the
orbital period. Filled dots refer to CORAVEL data, and open circles to 
radial velocities derived from photographic spectra obtained at the
Haute-Provence Observatory (1.52-m telescope). 
Measurements from Brown
et al. (1990) and from Beavers \& Eitter (1976) are represented by
open squares and open triangles, respectively.
Only the CORAVEL data, plus one OHP photographic measurement (open
circle with an error bar), have been used to derive the orbital solution
}
\end{figure} 

The radial-velocity curve, folded with the orbital period, is
presented in Fig.~\ref{Fig:VrHR363}. This figure presents as well
the other measurements 
(namely the three radial velocities from the Fick Observatory, 
the 15 measurements from Brown et al., and the 6 radial velocities we
obtained from photographic spectra at OHP) 
that were not used in the orbital solution,
since they would degrade its accuracy. These observations are nevertheless
compatible with the computed solution.
Note that the $O-C$ residuals 
are significantly larger than the accuracy of the measurements.
This jitter is likely due to envelope pulsations or to atmospheric
motions, as discussed by Jorissen et al. (1998; see their Fig.~1).

As for HD~191226, the mass function is compatible with a WD
companion,
since $M_1 = 1.5\pm0.5$~\Msun\ for the red giant implies $M_2 >
0.70\pm0.15$ \Msun\ for the unseen companion, given $\sin i \le
1.0$. If the companion is to be a WD with a mass typical of field WD's
(0.58~\Msun; Reid 1996), the orbital inclination has to be close to 
90$^\circ$, and the system may be an eclipsing  binary.

This star has been observed with the {\it International Ultraviolet
Explorer} (IUE) and ROSAT satellites. Although there is only
marginal
evidence for an UV continuum from a hot companion (Ake et al.
1988),
HR 363 is a strong source of hard X-rays (Jorissen et al. 1996). 
These X-rays are not expected to come from a hot corona, because
with $B-V = 1.7$, HR 363 lies far to
the right of the region of the Hertzsprung-Russell diagram
populated
by class III giants with a hot corona (H\"unsch et al. 1996).
The hard X-rays in HR 363 are therefore likely powered by
mass transfer in the binary system. The same holds true for the
weak
HeI $\lambda 1083.0$~nm emission line observed in HR 363 (Brown et al.
1990), since that line is generally absent in cool M giants, but is
frequent in interacting binary systems like symbiotic stars (Brown
et
al. 1990 and references therein).
It is somewhat surprising, though, that HR 363 behaves as an
interacting binary system while having the longest known period
(4593~d = 12.6~y) among S stars (Jorissen et al. 1998).
That question is discussed in more details in Sect.~\ref{Sect:UV}.

\section{Binary interaction vs. orbital separation}
\label{Sect:UV}

\tabcolsep 4pt
\begin{table*}
\caption[]{\label{Tab:symbiotic}
Signatures of binary interaction in extrinsic S stars, arranged in
order of increasing orbital period.  The orbital separations at
apastron and periastron ($A_{\rm apa}$ and $A_{\rm peri}$,
respectively) have been computed from Kepler's third law assuming
a total mass of 2 \Msun\ for all systems. In column `UV emission
lines': `no' = no emission lines detected; `IB' = emission lines
typical of interacting binaries; `WD' = WD continuum  
}
\begin{tabular}{llrlclclllc}
\hline
Name & Sp. Typ. & $P$     & \multicolumn{1}{c}{$e$} & $A_{\rm peri}-A_{\rm
apa}$ & \Mbol  
              & $L_{\rm X}$ & \multicolumn{2}{c}{UV} & H$_\alpha$&
HeI\cr
\cline{8-9}
     &          &  (d)   &     & (AU) &     & (\Lsun)
     & $L_{\rm SWP}$/\Lsun &  em. lines & em. &
$\lambda 1083.0$~nm \cr
\hline\cr
HD 121447& K4IIIBa4.5 & 185.7&\multicolumn{1}{c}{0.0}& $0.8-0.8$ & -1.4  & ?  
& $<0.002$      & no  & ? & ?\cr 
HR 1105  & S3.5/2     & 596.2       & $0.09\pm0.02$  & $1.6-1.9$ & -3.1  & ?  
& 0.003 - 0.02  & IB  & ? & var\cr
HD 35155 & S3/2       & 640.5       & $0.07\pm0.02$  & $1.7-2.0$ & -3.1$^z$
&
$<6$(-4) - 1(-3)$^x$ & 0.08 - 0.2 &IB&y &var\cr 
HD 191226& M1-3S      & 1210.4      & $0.19\pm0.02$  & $2.3-3.3$ & -$^z$ & ?  
& -             & WD  & ? & wk em\cr
HD 49368$^y$& S3/2    & 2995.9   & $0.36\pm0.05$  & $3.3-7.0$ & -3.4  & ?  
& 0.005 &  IB & ? & abs \cr
HR 363   & S3/2       & 4592.7      & $0.17\pm0.03$  & $5.7-8.0$ & -3.2  &
$<7$(-5) - 1.5(-3) & 0.002 &  no & ? & wk em
\medskip\cr
Ref      & a & \multicolumn{1}{c}{b}&
\multicolumn{1}{c}{b} & b  & c 
& d & e& f  
& g         & h\cr
\hline
\mbox{}\cr
\end{tabular}

References: a: Keenan \& Boeshaar (1980), Keenan \& McNeil (1989); 
b: Jorissen et al. (1998); c: Van Eck et al. (1998); d:
Jorissen
et al. (1996); e: Flux densities at 160 nm from Johnson et al. (1993)
combined with distances from 
Van Eck et al. (1998) or Eggen (1972); f: Johnson et al. (1993),
Ake (1997); g: Ake et al. (1991);  
h: Brown et al. (1990), Shcherbakov \& Tuominen (1992)
\smallskip\\
Remarks: x: $1(-3)$ stands for $1\; 10^{-3}$; 
y: HD 49368 = V613 Mon; z: HIPPARCOS parallax very uncertain; for
HD~35155, distance from Eggen (1972) adopted instead
\end{table*}

Binary S stars share many properties with symbiotic stars, since
both families consist of a cool red giant and a WD companion in
systems
with orbital periods of a few hundred to a few thousand days
(Jorissen
1997). Some kind of symbiotic activity should thus be expected
among
binary S stars as well.
Table~\ref{Tab:symbiotic} lists those systems where the usual
signatures of
binary interaction have been probed.
In that table, $L_{\rm X}$ refers to the luminosity in the
ROSAT hard band (0.5 -- 2.4 keV) taken from Jorissen et al. (1996),
and adapted to the new HIPPARCOS distances (Van Eck et al.
1998). 
Hard X-rays have been observed in HR 363 and HD 35155, and appear
strongly variable. 
In the UV domain, several stars exhibit
strong emission lines of highly ionized species typical of
interacting binary systems 
(like CIV $\lambda 155.0$~nm). In the column labelled `UV em. lines', 
`IB' stands for
`interacting binary', `no' stands for no emission lines seen, and
`WD'
indicates that the UV spectrum fits that of a clean WD. 
The continuum UV luminosity in the 125.0 -- 195.0 nm band  is
often larger than would
be expected from an isolated WD [see column `$L_{\rm SWP}$']. In
Table~\ref{Tab:symbiotic}, $L_{\rm
SWP} = 4\pi d^2\;70\; f(160 {\rm nm}) $ , where the average flux
density $f(160 {\rm nm})$
in the 125.0 -- 195.0 nm domain is
taken from Johnson et al. (1993) and the distance $d$ from Van Eck
et al. (1998) or
Eggen (1972). The UV luminosity is often strongly variable, in which
case the different observed values are listed.
The HeI $\lambda$ 1083.0~nm triplet generally confirms the UV
diagnostics.
%This triplet is generally absent or very weak in 
%cool M giants, but is strong  in interacting binary systems like 
%symbiotic stars (e.g. Brown et al. 1990). 
The three binary S 
stars flagged as `interacting binaries' from their UV features are
also
those showing strong and variable HeI $\lambda 1083.0$~nm lines,
whereas
the two stars (HR 363 and HD 191226) with no UV emission lines
exhibit
only a weak HeI triplet in emission.

Two important conclusions may be drawn from the observations
summarized in Table~\ref{Tab:symbiotic}:\\ 
(i) {\it the level of binary interaction does not appear to be 
correlated with the orbital period}. HD 49368 (=V613 Mon) for
instance exhibits
a much higher level of activity than the shorter-period system HD
191226, whereas the shortest-period system HD 121447 does not show
any
sign of interaction at all. Moreover, the maximum X-ray luminosities of HR
363
and HD 35155 are comparable despite very different orbital
periods;\\
(ii) {\it the activity appears to be strongly variable.} 

Orbital modulation is likely a major cause of the activity
variations,
as shown by Shcherbakov \& Tuominen (1992) and Ake
et al. (1994) in the well-documented case of HR 1105. 
%A variable obscuration of the UV continuum, correlated with the
%appearance of UV iron absorption lines, has been reported for HD
%35155
%(Ake et al. 1991) and HD 49368 (= V613 Mon; Ake 1997), much like
%the ``iron
%curtain'' phenomenon that has been observed in several symbiotic
%stars (Sion et
%al. 1993; Shore et al. 1994).  
However, at a given phase, important cycle-to-cycle variations
remain
(Ake et al. observed a variation by a factor of $\sim 3$ in the
$\lambda
146.0$~nm flux of HR 1105 at phase 0.3 in two different orbital
cycles), 
suggesting the existence of yet another cause of (secular)
variability.

Various physical processes are able to produce hard photons
modulated
by the orbital motion in a binary system:\\ 
1. Heating of the red-giant hemisphere facing a hot WD;\\
2. Accretion-powered hot spot;\\
3. Stream of gas from the red-giant wind, heated when funneled
through the
inner Lagrangian point.

A strong sensitivity of the activity level to the binary separation
is expected in the first and second cases. In the first case, the
orbital separation directly controls the dilution suffered by the hot
radiation when it reaches the giant atmosphere. In the second case, 
the mass accretion rate by the secondary roughly scales as
$\dot{M} k^4/(1+k^2)^{3/2}$ (where $\dot{M}$ is the wind mass-loss
rate of the giant, and $k$ is the ratio between the orbital and the
wind
velocities) in the case
of supersonic Bondi-Hoyle accretion in a detached system (see
Theuns et
al. 1996). The previous relation reduces to $\dot{M} A^{-2}$
(where $A$ is the orbital separation) when $k << 1$ (i.e. $v_{\rm
wind} >> v_{\rm orb}$).

From a detailed analysis of the variations with orbital phase of
the UV flux level and the HeI $\lambda 1083.0$~nm line shape, 
Shcherbakov \& Tuominen (1992) and Ake et al. (1994)
favour the third process as the origin of the hard photons,
i.e. funneling of the red-giant wind through the inner Lagrangian
point. 
The existence of such funneled streams is moreover predicted by  
{\it smooth particle hydrodynamics} simulations of mass transfer in
detached binary systems (Theuns \& Jorissen 1993; Theuns et al.
1996).
Different viewing angles of the stream during the orbital cycle
account for
the orbital modulation, whereas long-term fluctuations of the
mass-loss rate account for the secular variations (like those observed
in Mira variables, and associated with a clumpy and  non-spherically
symmetric wind; e.g. Whitelock et al. 1997, Lopez et al. 1997,
Olofsson 1997). 
The wind mass-loss rate of the red giant, rather than the orbital
separation, is expected to be the dominant factor controlling the
activity level in this case.  
The absence of any correlation between the
orbital periods and the activity levels in the sample
of S stars listed in Table~\ref{Tab:symbiotic} therefore suggests
that
streams like the one observed in the system HR 1105 might in fact
be responsible for the activity observed in other S stars as well. 
The absence of any activity observed in the system HD 121447,
despite
the fact that it is the closest system in the sample, may then be
attributed to its low luminosity (\Mbol $=-1.4$), and therefore
low mass-loss rate. 
Among the more luminous S stars (\Mbol $\sim -3.2$), differences in
their
mass-loss rates may account for their different activity  levels
(compare e.g. HD 35155 and HR 1105 having different activity levels
despite similar periods and spectral types, 
or HD 35155 and HR 363 having the same X-ray flux at
very different orbital periods). 

Future detailed studies of this class
of mass-losing, binary red giants may thus be expected to shed light on the
mass-loss
process, as well as on the physics of interacting binaries. 

\acknowledgements{We wish to express our thanks to Tom Ake for
communicating us results in advance of publication. Data and
bibliographic references made available by the {\it Centre de
Donn\'ees Stellaires} (Strasbourg) were of great help in the
present study. This work was supported in part by the {\it Fonds
National de la
Recherche Scientifique} (Belgium, Switzerland).}


\begin{thebibliography}{}

\bibitem[]{}
Abt H.A., 1970, ApJS 19, 387

\bibitem[]{}
Abt H.A., 1973, ApJS 26, 472

\bibitem[]{}
Ake T.B., 1997. In: Wing R.F. (ed.) The Carbon Star Phenomenon (IAU
Symp. 177). Kluwer, Dordrecht, in press

\bibitem[]{} 
Ake T.B., Johnson H.R., Peery B.F.Jr., 1988. In: Rolfe E.J. (ed.)
A 
Decade of UV Astronomy with IUE.  ESA-SP 281, p.245

\bibitem[]{}
Ake T.B., Johnson H.R., Ameen M.M., 1991, ApJ 383, 842

\bibitem[]{}
Ake T.B., Johnson H.R., Bopp B.W., 1994. In: Shafter A.W. (ed.)
Interacting Binary Stars. ASP Conf. Ser. 56, p. 413

\bibitem[]{}
Baranne A., Mayor M., Poncet J.L., 1979, Vistas in Astron. 23, 279

\bibitem[]{}
Barbier M., 1962, J. Observateurs 45, 57

\bibitem[]{}
Beavers W.I., Eitter J.J., 1986, ApJS 62, 147

\bibitem[]{}
Brown J.A., Smith V.V., Lambert D.L., Dutchover E.Jr., Hinkle K.H.,
Johnson H.R., 1990, AJ 99, 1930

\bibitem[]{}
Carquillat J.M., Jaschek C., Jaschek M., Ginestet N., 1997, A\&AS 123,
5

\bibitem[]{}
Eggen O.J., 1972, ApJ 177, 489
 
\bibitem[]{}
Ginestet N., Carquillat J.M., Jaschek M., Jaschek C., 1994, A\&AS
108,
359 

\bibitem[]{}
Harper W.E., 1934, Publ. Dominion Astrophys. Obs. 6, 151

\bibitem[]{}
H\"unsch M., Schmitt J.H.M.M., Schr\"oder K.-P., Reimers D., 1996,
A\&A
310, 801

\bibitem[]{}
Jaschek C., Jaschek M., 1987. The Classification of Stars,
Cambridge
Univ. Press

%\bibitem[]{}
%Johnson H.R., Ake T.B., 1989. In: Johnson H.R., Zuckerman B.
%(eds.)
%Evolution of Peculiar Red Giants (IAU Coll. 106). Cambridge U.P.,
%p. 371

\bibitem[]{}
Johnson H.R., Ake T.B., Ameen M.M., 1993, ApJ 402, 667

\bibitem[]{}
Jorissen A., 1997. In: Mikolajewska J. (ed.) Physical Processes in
Symbiotic Stars  and Related Systems. 
Copernicus Found. for Polish Astronomy, Warsaw, p.135

\bibitem[]{}
Jorissen A., Schmitt J.H.M.M., Carquillat J.M., Ginestet N.,
Bickert
K.F., 1996, A\&A 306, 467

\bibitem[]{}
Jorissen A., Van Eck S., Mayor M., Udry S., 1998, A\&A, in press

\bibitem {}
Keenan P.C., 1954, ApJ 120, 484

\bibitem {}
Keenan P.C., Boeshaar P.C., 1980, ApJS 43, 379

\bibitem {}
Keenan P.C., McNeil R.C., 1989, ApJS 71, 245

\bibitem[]{} 
Little, S.J., Little-Marenin, I.R., Hagen-Bauer, W., 1987, AJ
94, 981 

\bibitem[]{} 
Lopez B., Danchi W.C., Bester M., et al., 1997, ApJ, in press

\bibitem[]{}
Nadal R., Ginestet N., Carquillat J.M., P\'edoussaut A., 1979,
A\&AS
35, 203

\bibitem[]{}
Nassau J.J., McRae D.A., 1949, ApJ 110, 478

\bibitem[]{}
Olofsson H., 1997. In:
Wing R. (ed.) The Carbon Star Phenomenon (IAU
Symp. 177). Kluwer, Dordrecht, in press

\bibitem[]{}
Reid I.N., 1996, AJ 111, 2000

\bibitem[]{}
Sato K., Kuji S., 1990, A\&AS 85, 1069

\bibitem[]{}
Schmidt-Kaler Th., 1982. In: Landolt-B\"ornstein (ed.) Zahlenwerte
und
Funktionen aus Naturwissenschaften und Technik, Group 6,
Vol. 2. Springer-Verlag, Berlin, p. 1

\bibitem[]{}
Shcherbakov A.G., Tuominen I., 1992, A\&A 255, 215

\bibitem[]{}
Smith V.V., Lambert D.L., 1988, ApJ 333, 219

\bibitem[]{}
Theuns T., Jorissen A., 1993, MNRAS 265, 946

\bibitem[]{}
Theuns T., Boffin H.M.J., Jorissen A., 1996, MNRAS 280, 1264

\bibitem[]{}
Udry S., Mayor M., Van Eck S., Jorissen A., 1998a, A\&AS, in press

\bibitem[]{}
Udry S., Mayor M., Van Eck S., Jorissen A., Pr\'evot L., Grenier S., Lindgren
H., 1998b, A\&AS, in press

\bibitem[]{}
Van Eck S., Jorissen A., Udry S., Mayor M., Pernier B., 1998, A\&A 329, 971

\bibitem[]{}
Whitelock P.A., Feast M.W., Marang F., Overbeek M.D., 1997, MNRAS 288,
512

\bibitem[]{}
Wilson R.E., 1953. General Catalogue of Stellar Radial Velocities, 
Carnegie Inst. Washington Publ. 601 (GCSRV)

\bibitem[]{}
Wilson R.E., Joy A.H., 1952, ApJ 115, 157

\bibitem[]{}
Yamashita Y., 1967, Publ. DAO 13, 47

\bibitem[]{}
Young R.K., 1945, Publ. DDO 1, 311

\end{thebibliography}
\end{document}